\documentclass[twocolumn,showpacs,preprintnumbers,amsmath,amssym,article]{revtex4}
\usepackage{mathrsfs}
\usepackage{dcolumn}
\usepackage{bm}
\usepackage{graphicx}
\usepackage{MnSymbol,wasysym}
\begin{document}

\title{Braiding a novel kind of Majorana-like quasiparticles in nanowire quantum dots}
\author{Kuo Hai, \ Qiong Chen, \ Wenhua Hai}

\affiliation{Department of Physics and Key Laboratory of Low-dimensional Quantum Structures and Quantum Control of Ministry
of Education, and Synergetic Innovation Center for Quantum Effects and Applications, Hunan Normal University, Changsha 410081, China}
\email{ron.khai@gmail.com; whhai2005@aliyun.com}

\begin{abstract}
For an electrically driven electron confined in a nanowire quantum dot with spin-orbit coupling (SOC), we find a SOC-magnetism phase-locked condition under which we derive a complete set of Schr\"odinger kitten states which contains some novel degenerate ground states with oscillating wave packets or stationary double packets in undriven case. We identify such wave packets as Majorana-like quasiparticles and demonstrate that they obey non-Abelian statistics and behave similarly to neutral particles. The braiding operations based on the interchanges of the degenerate non-Abelian quasiparticles are shown, which shift the system between different ground states and may be insensitive to perturbations and weak noise from the environment. The results could be tested experimentally in the existing setups and could be treated as the leading-order results to directly extended to an array of weakly coupled single-electron quantum dots for encoding topological qubits.
\end{abstract}


\maketitle

\section{Introduction}

The spin-orbit coupling (SOC) describes interaction between
the motion of an electron and its spin. For a single electron, the SOC can hybridize spin-up and spin-down states form a spin-orbit qubit \cite{Nowack,NPerge2,Ladriere,Mourik,NPerge3,RLi}. The orbital part of the spin-orbit wavefunction which is similar to the spin-orbit entangled states in different systems \cite{Leibfried,Monroe,Kitagawa} can be used for qubit manipulation. Coherent manipulation of electron spin is one of the central problems of
spintronics \cite{Wolf,Loss,Golovach} and is of critical importance for quantum computing and
information processing with spins \cite{Loss}. An interesting proposal \cite{Loss} suggested that the spin of an electron confined to a
quantum dot can be used as a basic qubit to store and process quantum information, and the results were extended to an array of quantum dots which was operated by a set of quantum gates that act on single spins and pairs of neighboring spins. The previous investigation has paved the way for manipulating
electron spins in quantum dots individually \cite{Kato,Rashba}. Recently, the search for non-Abelian quasiparticles in semiconducting nanowire and quantum dot with strong SOC has been a focus of theoretical and experimental efforts \cite{Elliott,Aguado,Mourik,Das,Deng,Ptok,Nilsson,Gazibegovic}, motivated by their potential utility for fault tolerant topological quantum computation \cite{Kitaev,Nayak,Stern,Moore,DSau,Stern2,YLi,Else}.

A Majorana particle being in ground states  \cite{Majorana,Kitaev,LFu, Gazibegovic} is an electrically neutral non-Abelian anyon \cite{Moore,Stern,Abdumalikov,Wilcze} identical to its own antiparticle. Interchanging the Majorana particles changes the state of the system in a way that depends only on the order in which the exchange was performed, which is the base of braiding operations for topological quantum computation \cite{Stern}. Although no sightings of a Majorana particle have been reported in the elementary particle world, its existence has been demonstrated as particle-like excitations called quasiparticles and can be used to form the Majorana bound states in condensed matter physics.
There are some candidate sources of the Majorana quasiparticles such as the fractional $5/2$ quantum Hall state \cite{Moore} carries one-quarter of an electron charge, the semiconductor nanowires in contact with a superconductor \cite{Mourik,NPerge3,Aasen}, Shockley states at the end points of superconducting wires or line defects \cite{Kitaev}, the quantum vortex in certain two-dimensional superconductors or superfluids \cite{Kopnin} and the spin-polarized resonant level in the vicinity of the quantum critical point \cite{Mebrahtu}. It has also been demonstrated theoretically \cite{Lutchyn,Oreg} and experimentally \cite{Mourik,NPerge,QHe} that the elusive Majorana particles can be detected in some one-dimensional (1D) systems, including the semiconducting nanowire quantum dot with strong SOC and large $g$ factor \cite{Mourik,Nilsson}, and in proximity to a superconductor. The non-Abelian statistics of Majorana bound states and their controllable entanglement allow them to be used in carrying out topological quantum computation \cite{Kitaev,Nayak,DSau}. High $g$ factors and strong SOC, and the ability to induce superconductivity put forward InSb nanowires as a natural platform for the  realization of 1D topological states \cite{NPerge3}. The key braiding operation of non-Abelian anyons has been implemented by using 1D semiconducting wire networks by adjusting gate voltages \cite{Alicea}.
The search for Majorana fermions in 1D conductors is focused on finding the best material in terms of a strong
spin-orbit interaction and large Lande¡ä $g$ factors \cite{NPerge3}. One of the current main objectives may be the investigation of novel models for finding non-Abelian quasiparticles \cite{Shtengel,Feldman}.

Mathematically, a partially differential system allows a general solution with arbitrary functions and a complete solution with arbitrary constants. These arbitrary functions and constants are adjusted and determined by the initial and boundary conditions. The general solution can describe all properties of the system and the complete solution can also describe more physics than any particular solution can. In the previous work, we derived a set of generalized coherent states for a trapped ion system \cite{Hai,Hai2}, which just is a set of complete solutions describing a complete set of Schr\"odinger kitten (or cat) states. As pointed out in Ref. \cite{Ourjoumtsev}, ``a Schr\"odinger kitten (cat) state is usually defined as a quantum superposition of coherent states with small (big) amplitudes. The amplitude of the coherent states can be amplified to transform the Schro\"odinger kittens into
bigger Schr\"odinger cats, providing an essential tool for quantum information processing." For an ion system the similar spin-motion entangled states have been experimentally prepared as the Schr\"odinger's cat state with two macroscopically separated wave packets \cite{Monroe,Kienzler}. Spin-orbit qubit for an electron system in a semiconductor nanowire has also been investigated, by using SOC which provides a way to control spins electrically \cite{NPerge2,Kato,Rashba}.
Here we are interested in how the wave packets described by the generalized coherent states replace the vortices \cite{Stern,Kopnin} as the Majorana-like quasiparticles. \emph{Such quasiparticles behave as electroneutrality without Coulomb interaction between them and their interchange in one spatial dimension becomes possible with one wavepacket going through another.}

In this paper, we consider a spin-orbit coupled and electrically driven electron confined in a nanowire quantum dot. We find that when the orientation of the static magnetic field and SOC-dependent phase fits a \emph{SOC-magnetism phase-locked condition}, the system has a set of complete solutions of Schr\"odinger equation with arbitrary constants adjusted by the initial conditions, which describes a complete set of Schr\"odinger kitten states and contains some degenerate ground states with novel oscillating wave packets. For the undriven case and in the \emph{magnetic resonance} case, stationary double packets of degenerate ground states are constructed. The degeneracy is not based on simple symmetry consideration and is topological thereby. We identify such wave packets as Majorana-like quasiparticles and demonstrate that they obey non-Abelian statistics and behave as electroneutrality without Coulomb interaction between them. The braiding operations based on the interchanges of the degenerate non-Abelian quasiparticles with one wavepacket going through another are shown, which shift the system between different ground states and may be insensitive to perturbations and noise from the environment. Based on the exact solutions, the results could be tested experimentally in the existing setups and could be directly extended to an array of electrons separated from each other by different 2D quantum dots with weak neighboring coupling \cite{Loss} for topological quantum computation.

\section{A complete set of Majorana-like degenerate ground states}

We consider a gated nanowire quantum dot with Rashba-Dresselhaus coexisted SOC,
where a single electron is confined in a 1D harmonic well controlled by the gate
voltages on the static electric gates, and
subject to an arbitrarily strong static magnetic field \cite{Pershin,Nowak} and an arbitrarily strong ac electric field. The Hamiltonian governing the system reads \cite{RLi}
\begin{eqnarray}\label{dy3}
H&=&H_0+\alpha_D\sigma_x p+\alpha_R\sigma_y p+\frac 1 2 g_0\sigma_n \nonumber\\
H_0&=&-\frac{1}{2}\frac{\partial^2}{\partial x^2}+\frac 1 2 x^2+\epsilon x\cos(\Omega t),\nonumber\\
g_0&=&g_e \mu_B B,\ \ \sigma_n=\sigma_x\cos \theta+\sigma_y\sin \theta,
\end{eqnarray}
where we have adopted the natural unit system with $\hbar=m_e=\omega=1$ so that time, space and energy are in units of $\omega^{-1},\ L_h=\sqrt{\hbar/(m_e\omega)}$ and $\hbar\omega$. Here $m_e$ is the effective electron mass, $\omega$ denotes the trapped frequency, $\alpha_{R(D)}$ is the Rashba (Dresselhaus) SOC strength, $\sigma_{x(y)}$ is the $x (y)$ component of Pauli matrix, $g_e$ denotes the gyromagnetic ratio \cite{Tsitsishvili}, $\mu_B$ is the Bohr magneton, $B$ and $\theta$ represent the strength and orientation of the static magnetic field, and $\epsilon$ and $\Omega$ are the amplitude and frequency of the ac electric field. Applying the usual state vector $|\psi(t)\rangle=\frac{1}{\sqrt 2}[|\psi_+ (t)\rangle|\uparrow \rangle+|\psi_- (t)\rangle|\downarrow \rangle]$, the space-dependent state vector is defined as
\begin{eqnarray}\label{dy3}
|\psi(x,t)\rangle=\langle x|\psi(t)\rangle=\frac{1}{\sqrt 2}\Big[\psi_+ (x,t)|\uparrow \rangle+\psi_- (x,t)|\downarrow \rangle\Big]
\end{eqnarray}
with $\psi_{\pm}(x,t)=\langle x|\psi_{\pm}(t)\rangle$ being the normalized motional states entangling the corresponding spin states $|\uparrow \rangle=\left(\begin{array}{c}
  1 \\ 0
\end{array}\right)$ and $|\downarrow \rangle=\left(\begin{array}{cc}
  0 \\ 1
\end{array}\right)$, respectively, where $|\psi_{\pm}(t)\rangle$ may be expanded in terms of a set of orthonormal basic kets with time-dependent expansion coefficients \cite{Leibfried}. However, here we will seek the exact complete solutions. Therefore, the spin-orbit entanglement of Eq. (2) requires the linear independencies \cite{Hai2,Kong} of $\psi_+(x,t)$ and $\psi_-(x,t)$.
The probabilities of the particle being in spin states $|\uparrow \rangle$ and $|\downarrow \rangle$ are $P_{\pm}(t)=\frac 1 2 \int|\psi_{\pm} (x,t)|^2dx$. The maximal spin-orbit entanglement can be associated with \cite{Hai2} $P_+=P_-=\frac 1 2$. Applying Eqs. (1) and (2) to the Schr\"odinger quation $i\frac{\partial|\psi(x,t)\rangle}{\partial t}=H|\psi(x,t)\rangle$ yields the matrix equation
\begin{widetext}
\begin{eqnarray}
i\frac{\partial}{\partial t}\left(\begin{array}{c}  \psi_+ \\ \psi_- \end{array}\right)=H_0\left(\begin{array}{c}
   \psi_+ \\  \psi_- \end{array}\right)-i\alpha \frac{\partial}{\partial x}\left(\begin{array}{c}
    e^{-i\phi}\psi_- \\  e^{i\phi} \psi_+ \end{array}\right)+\frac 1 2 g_0 \left(\begin{array}{c}  e^{-i\theta}\psi_- \\ e^{i\theta}\psi_+ \end{array}\right)
    \ \ \ \text{for}  \ \ \alpha=\sqrt{\alpha_D^2+\alpha_R^2},\ \phi=\arctan \frac{\alpha_R}{\alpha_D},
\end{eqnarray}
where we have taken the definitions of the SOC strength and SOC-dependent phase as \cite{RLi} $\alpha$ and $\phi$ for the Rashba-Dresselhaus SOC coexistence system. Making the function transformations
\begin{eqnarray}\label{dy3}
\psi_{\pm}(x,t)=\psi_{\pm, l}(x,t)=e^{\mp i\phi/2} [u(x,t)e^{-i(\alpha x+l\pi/2)}\pm v(x,t)e^{i(\alpha x+l\pi/2)}]\ \ \ \text{for}\ \ \ l=0,1,...,
\end{eqnarray}
and inserting it into Eq. (3), then multiplying the first line of the matrix equation by $e^{i(\phi/2+\alpha x+l\pi/2)}$ and multiplying the second line of the equation by $e^{-i(\phi/2+\alpha x+l\pi/2)}$, we obtain
\begin{eqnarray}
i\frac{\partial}{\partial t}\left(\begin{array}{c} u \\ v \end{array}\right)=\Big(H_0-\frac{\alpha^2}{2}\Big)\left(\begin{array}{c}
   u \\  v \end{array}\right)+ \frac 1 2 g_0 \left(\begin{array}{c}  e^{-i(\theta-\phi)}[u-v e^{i (2\alpha x+l\pi)}]+e^{i(\theta-\phi)}[u+v e^{i (2\alpha x+l\pi)}] \\ e^{-i(\theta-\phi)}[u e^{-i (2\alpha x+l\pi)}-v]-e^{i(\theta-\phi)}[u e^{-i (2\alpha x+l\pi)}+v] \end{array}\right).
\end{eqnarray}
\end{widetext}
For an arbitrary angle $\theta$, the final term  of  Eq. (5) cannot be decoupled, so it is hard to construct an exact solution of the system. The corresponding perturbed solution has been considered in Ref. \cite{RLi} that leads to some interesting results. The sensitivity of exact solution to the magnitude and direction of applied magnetic fields is in good agreement with experimental observation \cite{Mourik} and matches theoretical expectation for the states associated with Majorana quasiparticles \cite{Wilcze}. Here we are interested in the \textbf{SOC-magnetism phase-locked} case $\theta=\phi=\arctan \frac{\alpha_R}{\alpha_D}\to \phi_0+ j\pi$ for $j=0,1,2,...$ and $\phi_0\in[0, \frac \pi 2]$,
which can be realized experimentally for any fixed SOC strengths $\alpha_R$ and $\alpha_D$ by selecting the proper orientation of magnetic field, because of the multivaluedness of inverse tangent function. Under such a condition, Eq. (5) becomes the decoupled equation
\begin{eqnarray}
i\frac{\partial}{\partial t}\left(\begin{array}{c} u \\ v \end{array}\right)=\Big[H_0(x,t)-\frac{\alpha^2}{2}+ g_0\sigma_z \Big]\left(\begin{array}{c}
   u \\  v \end{array}\right) \ \ \text{for} \ \ \theta=\phi,
\end{eqnarray}
where $\sigma_z$ is $z$ component of the Pauli matrix. Equation (6) can be regarded as a new two-level system of the effective Hamiltonian $H_{eff}=H_0-\frac{\alpha^2}{2}+ g_0\sigma_z$.

After making the new function transformations
$u= \frac {c_u}{\sqrt 2} f_u(x,t)e^{i(\alpha^2/2-g_0)t}, \  v= \frac {c_v}{\sqrt 2 } f_v(x,t)e^{i(\alpha^2/2+g_0)t}$ with $c_u$ and $c_v$ being the complex constants determined by the normalization and initial conditions,
the decoupled Eq. (6) gives the time-dependent Schr\"odinger quation $i\frac{\partial f_{u(v)}}{\partial t}=H_0 f_{u(v)}$ of a driven harmonic oscillator with the exact complete solutions being the orthonormal generalized coherent states \cite{Hai,Hai2}
\begin{eqnarray}
f_{u(v)}&=&f_{n_{u(v)}}=R_{n_{u(v)}}(x,t)e^{i\Theta_{n_{u(v)}}(x,t)},  \nonumber \\
\Theta_{n_{u(v)}}&=&-\Big(\frac{1}{2}+n_{u(v)}\Big)\chi(t)+b_{u(v) 2}x+\frac{\dot{\rho}}
{2\rho}x^2+\gamma_{u(v)}(t),  \nonumber \\
R_{n_{u(v)}}&=&\Big(\frac{\sqrt{c_0}}{\sqrt{\pi}2^{n_{u(v)}}n_{u(v)}!\rho}\Big)^{\frac{1}{2}}H_{n_{u(v)}}[\xi_{u(v)}]e^{-\frac{1}{2}
\xi^2_{u(v)}},\ \ \
\end{eqnarray}
for $n_u, n_v=0,1,2,...$ with $R_{n_{u(v)}}(x,t)$ and $\Theta_{n_{u(v)}}(x,t)$ being the real functions and $H_{n_{u(v)}}[\xi_{u(v)}]$ the Hermite polynomial of the space-time combined
variable $\xi_{u(v)}(x,t)=\frac{\sqrt{c_0}}{\rho(t)}x-\frac{b_{u(v) 1}(t)\rho(t)}{\sqrt{c_0}}$.
In Eq. (7), the real functions $\rho(t), \chi(t)$, $\gamma_{u(v)}(t),
b_{u(v) 1}(t)$ and $b_{u(v) 2}(t)$ have the forms
$\rho(t)=\sqrt{\varphi_1^2+\varphi_2^2}$, $\chi(t)=\arctan\Big(\frac{\varphi_2}{\varphi_1}\Big)$,
$b_{u(v) 1}(t)=\frac{\epsilon}{\rho^2(t)}[\varphi_1(t) \int_0^t \varphi_2(\tau)\cos(\Omega \tau)d \tau -\varphi_2 (t)\int_0^t \varphi_1(\tau)\cos(\Omega \tau)d \tau]+b_{u(v) 1}(0)\varphi_1(t)+b_{u(v) 2}(0)\varphi_2(t)$, $b_{u(v) 2}(t)=\frac{\epsilon}{\rho^2(t)}[-\varphi_1(t) \int_0^t \varphi_1(\tau)\cos(\Omega \tau)d \tau -\varphi_2(t) \int_0^t \varphi_2(\tau)\cos(\Omega \tau)d \tau]+b_{u(v) 2}(0)\varphi_1(t)-b_{u(v) 1}(0)\varphi_2(t), \gamma_{u(v)}(t)=\frac{1}{2}\int_0^t[b_{u(v) 1}^2(\tau)-b_{u(v) 2}^2(\tau)]d\tau+\gamma_{u(v)}(0)$.
Here $\varphi_{1,2}(t)=A_{1,2} \cos(t+B_{1,2})$ are the real and imaginary parts of the solution to equation $\ddot \varphi =-\varphi$ with oscillating frequency $1(\omega)$, $A_{1,2}, B_{1,2}$ and $c_0=\varphi_1\dot\varphi_2-\varphi_2\dot\varphi_1=A_1 A_2\sin(B_1-B_2)$ are the initial
constants. The initial constant sets $S_{u(v)}=[\gamma_{u(v)}(0)], b_{u(v) 1}(0), b_{u(v) 2}(0),A_{1,2}, B_{1,2}]$ are determined by the form of the initial states \cite{Hai2} and the initial coherent states can be experimentally prepared  \cite{Monroe,Ourjoumtsev}. Then the solutions $f_{u(v)}=f_{n_u(n_v)}[S_{u(v)},x,t]$ are determined by the sets $S_{u(v)}$ for fixed quantum numbers $n_{u(v)}$.

Applications of $u$ and $v$ to Eq. (4) result in new forms of the exact complete solutions of Eq. (3) as
\begin{eqnarray}
\psi_{\pm, l n_u n_v}(x,t)&=& \frac {1}{\sqrt 2} e^{\frac i 2(\alpha^2 t\mp \phi-l\pi)} [c_u f_{n_u}e^{-i(\alpha x+g_0 t)} \nonumber \\ &&\pm c_v f_{n_v}e^{i(\alpha x+g_0 t+l\pi)}].
\end{eqnarray}
In Eq. (8), the quantum numbers $l,n_{u(v)}$ are independent of the parameters in system (1). The solutions $f_{n_u(n_v)}$ of Eq. (7) can be the eigenstates of a harmonic oscillator for the undriven case with $\epsilon=0$ and the generalized coherent states for any driving strength \cite{Hai,Hai2}, which lead to different forms of Eq. (8) and the corresponding rich physics. Clearly, for any nonzero function pair $f_{u(v)}$ and nonzero constants $\alpha, g_0$, the solutions $\psi_{+, l n_u n_v}(x,t)$ and $\psi_{-, l n_u n_v}(x,t)$ are linearly independent, so the superposition state (2) is spin-orbit entangled. It is important to note that in Eqs. (7) and (8), $f_{n_u(n_v)}$ depends only on the ac driving and trapping field, and is independent of the SOC, the static magnetic field and the quantum number $l$. Therefore, we can conveniently manipulate the quantum states (8) by independently adjusting the driving and the initial constants to select the exact solutions of Eq. (7), and by independently tuning the SOC parameters $\alpha, \phi$ and magnetic field parameter $g_0$.  Applying Eq. (8) to Eq. (2) and noticing $\psi_{\pm, l n_u n_v} (x,t)=\langle x|\psi_{\pm, l n_u n_v}\rangle$, we arrive at the orthonormal complete set of the exact superposition states,
\begin{eqnarray}
|\psi_{l n_u n_v}\rangle=\frac{1}{\sqrt 2}[|\psi_{+, l n_u n_v}\rangle|\uparrow \rangle+|\psi_{-, l n_u n_v}\rangle|\downarrow \rangle].
\end{eqnarray}
By making use of the orthonormalization of $f_{n_u(n_v)}$ and Eq. (7), the expected energy of the system reads \cite{Hai}
\begin{eqnarray}
&& E_{n_u n_v}(t)= i\langle\psi_{l n_u n_v}| \frac{\partial}{\partial t}|\psi_{l n_u n_v}\rangle
\nonumber\\
&=& -\frac{\alpha^2}{2}+ \frac i 2 \int_{-\infty}^{\infty}\Big(|c_u|^2f_{n_u}\frac{\partial f_{n_u}}{\partial t}+|c_v|^2f_{n_v}\frac{\partial f_{n_v}}{\partial t}\Big)dx \nonumber\\
&=&-\frac{\alpha^2}{2}- \frac 1 2 \int_{-\infty}^{\infty}\Big(\dot{\Theta}_{n_{u}}|c_u|^2R_{n_u}^2+\dot{\Theta}_{n_{v}}|c_v|^2R_{n_v}^2\Big)dx.\ \ \ \ \ \
\end{eqnarray}
It is worth noting that the states of Eq. (8) depend on the magnetic field strength and number $l$ but the energy of Eq. (10) is independent of $g_0$ and $l$. Therefore, for a given $g_0$ and fixed quantum numbers $n_u, n_v$, different $l$ labels different degenerate states of Eq. (9)
with the density wave packets
\begin{eqnarray}
&&|\psi_{\pm, l n_u n_v}(x,t)|^2=\frac 1 2 [|c_u|^2R_{n_u}^2+|c_v|^2R_{n_v}^2] \pm D_{uv}, \nonumber\\
&&D_{uv}(x,t)=|c_u c_v| R_{n_u}R_{n_v}\nonumber\\
&&\times \cos [\Theta_{n_u}-\Theta_{n_v}-2(\alpha x+g_0 t)-l\pi+\phi'],\ \ \ \ \
\end{eqnarray}
where constant $\phi'$ is the phase difference, $\phi'=\arg c_v-\arg c_u$, the term $D_{uv}(x,t)$ describes the phase coherence with signs ``$\pm$" implying different coherent effects for the different motional states. The orthonormalization of Eq. (7) means that the probabilities of the particle being in spin states $|\uparrow \rangle$ and $|\downarrow \rangle$ obey $P_{+, l n_u n_v}(t)+P_{-, l n_u n_v}(t)=\frac 1 2 \int (|c_u|^2R_{n_u}^2+|c_v|^2R_{n_v}^2)dx=\frac 1 2 (|c_u|^2+|c_v|^2)=1$ which confines the normalization constants. Therefore, the probabilities of the particle occupying spin states $|\uparrow \rangle$ and $|\downarrow \rangle$ become $P_{\pm}(t)=\frac 1 2 \int|\psi_{\pm} (x,t)|^2dx=\frac 1 2 [1\pm \int D_{uv}(x,t)dx]$. Given $c_u, c_v$, $n_u$ and $n_v$, for a fixed $l$ Eq. (11) contains two different wave packets with the sign ``$+$" or ``$-$" respectively, while for a fixed sign of ``$\pm$" Eq. (11) also contains only a pair of different wave packets $|\psi_{\pm,l n_u n_v}|^2$ with even $l$ or odd $l$ respectively. Clearly, Eq. (11) shows $|\psi_{+,l n_u n_v}|^2=|\psi_{-,l' n_u n_v}|^2$ with $l$ and $l'$ possessing different odevity. Therefore, Eqs. (8) and (9) mean that $|\psi_{l n_u n_v}\rangle$ and $|\psi_{l' n_u n_v}\rangle$ are two different degenerate states associated with some interchanges of wave packets. The wave packets $|\psi_{+,l n_u n_v}|^2$ and $|\psi_{-,l n_u n_v}|^2$ may be spatially separated and centred at different positions, that means Eq. (9) being a complete set of electronic Schr\"odinger kitten states \cite{Ourjoumtsev} which includes the degenerate ground states with the smallest sum $n_u+ n_v$ of quantum numbers and different initial constants. We will take some simple cases to demonstrate that such degenerate ground states can be identified as Majorana-like quasiparticles governed by non-abelian statistics.

Let us extend the definition of a cat state at a selected time (e.g. the initial time) with the macroscopically separated wave packets \cite{Monroe} to the definition of a ``kitten state" with smaller maximal distance between two wave packets \cite{Ourjoumtsev}, where \emph{$|\uparrow \rangle$ and $|\downarrow \rangle$ refer to the internal states of an atom that has not and has radioactively decayed, while the right and left wave packets refer to the live $\smiley{}$ and dead $\frownie{}$ states of a kitten}. We can formally write a ground state of Eq. (9) as a Schr\"odinger kitten state with even $l$ as $|\psi_{l n_u n_v}\rangle=\frac{1}{\sqrt{2}}(|\smiley{}\rangle|\uparrow\rangle+|\frownie{}\rangle|\downarrow\rangle)$. Then its a degenerate ground state with odd $l'$ reads $|\psi_{l' n_u n_v}\rangle=\frac{1}{\sqrt{2}}(|\frownie{}\rangle|\uparrow\rangle+|\smiley{}\rangle|\downarrow\rangle)$ with exchange between wave packets $\smiley{}$ and $\frownie{}$ or equivalent spin flip, which is of a \textbf{``ill kitten"} transferring from near-dead to alive when the atom has radioactively decayed. For a group of fixed quantum numbers $l, n_u, n_v$ \emph{we identify the wave packets  $|\psi_{+, l n_u n_v}(x,t)|^2$ and  $|\psi_{-, l n_u n_v}(x,t)|^2$ as Majorana-like quasiparticle pair and will demonstrate that they obey non-Abelian statistics.} It is worth noting that the name ``Schr\"odinger kitten" of the superposition states is determined only by the spatially separated norms of the motional states at a selected time such that it can be related to many kitten states distinguished by the different phases. The braiding operations for topological quantum computation just are based on the transitions among such degenerate kitten states. The degeneracy of ground kitten states is not based on simple symmetry consideration and is topological thereby. Particularly, for some values of SOC strength the motional states of the kittens and ill kittens may exist zero-density nodes similar to the planar vortex cores \cite{Stern,Kopnin} at which the phases of states occur jumps with nonzero topological charges.

\section{Braiding the degenerate non-Abelian quasiparticles}

Firstly, let us take two simple examples with $|c_u|=|c_v|=1, n_u=0, n_v=0, 1$ and the initial constant set $S_{u(v)}=[\gamma_{u(v)}(0)=b_{u(v) 2}(0)=0, b_{u 1}(0)=-b_{v 1}(0)=b_0, A_1=A_2=A, B_1=0, B_2=-\pi/2]$ to demonstrate Non-Abelian statistics of the Schr\"odinger kittens. These constants are associated with $\varphi_1=A\cos t, \varphi_2=A\sin t, \rho=c_0=A, \chi =t, \xi_{u(v)}=x-b_{u(v)1}(t)$,
$b_{u 1}=\epsilon[\cos t \int_0^t \sin\tau\cos(\Omega \tau)d \tau -\sin t\int_0^t \cos\tau\cos(\Omega \tau)d \tau]+b_0\cos t=b_{v 1}+2b_0\cos t$, $b_{u 2}(t)=\epsilon[-\cos t \int_0^t \cos \tau\cos(\Omega \tau)d \tau -\sin t \int_0^t \sin\tau\cos(\Omega \tau)d \tau]-b_0\sin t=b_{v 2}-2b_0\sin t$ and $R_{0_{u}}(x,t)=\pi^{-1/4}e^{-\xi_{u}^2/2}$, $\Theta_{0_{u(v)}}(x,t)=-(\frac{1}{2}+n_{u(v)})t+b_{u(v) 2}x+\gamma_{u(v)}(t)$. Applying these constants and functions to Eqs. (7) and (8), we obtain the explicit solutions
\begin{eqnarray}\label{dy3}
&&\psi_{\pm, l 0 n_v}(x,t)=\frac{\pi^{-\frac 1 4}}{\sqrt 2} e^{\frac i 2[\alpha^2 t\mp \phi-l\pi-2(\alpha x+g_0 t)+2\Theta_{0_u}]}
\nonumber\\
&\times& \Big[e^{-\frac 1 2 \xi_u^2} \pm \frac{H_{n_v}(\xi_v)}{\sqrt{2^{n_v}n_v!}} e^{i(2\alpha x+2g_0 t+l\pi+\Theta_{0_v}-\Theta_{0_u}+\phi')-\frac 1 2 \xi_v^2}\Big].\ \ \ \ \
\end{eqnarray}
Here phase difference $\phi'$ between $c_v$ and $c_u$ is adjusted by the initial conditions governing the wave packets.

Combining Eq. (12) with Eq. (11), we display the spatiotemporal evolutions of the quasiparticle wave packets in Fig. 1 for the parameters $\alpha=0.2, \phi'=3, \epsilon =0.2, \Omega=2, n_u=0$ and (a, b) $g_0=0.1, b_0=1, n_v=0$; (c, d) $g_0=0.5, b_0=0, n_v=1$, where $l=0, 1$ is labelled as the subscripts of states. We can see the complicated spatiotemporal evolutions and the time periodicity as shown in Fig. 1(a). In Fig. 1(b) we show the ill kitten state at time $t=5.0(\omega^{-1})$ and kitten state at time $t=13.8(\omega^{-1})$ with wave packets interchange. In both case, the two separated peaks have a distance in order of $L_h=\sqrt{\hbar/(m_e\omega)}$. In the time interval $t\in (5.0, 13.8)$, there exist many pairs of wave packets with  different shapes. The same wave packets will periodically appear and the corresponding states may change their phases adjusted by the function $\Theta_{0_u}(x,t)$, which are related to the field intensities $\alpha, g_0$ and frequency $\Omega$.  Particularly, the interchanges of integer times make the norms of $\psi_{\pm, 0 0 1}(x,t)$ back to the original spatial distributions but their phases will evolve to different distributions. As a consequence, the quasiparticles described by the wave packets are known as Non-Abelian anyons. Thus we have demonstrated the Non-Abelian characteristic of the Schr\"odinger kitten states.

For the case $g_0=0.5, b_0=0, n_v=1$, the simple function relations $\xi_u=\xi_v, \Theta_{0_v}-\Theta_{0_u}=-2g_0t$ and $H_{n_v}(\xi_v)=2\xi(x,t)$ lead to simplification of Eq. (12). The asymmetrical spatiotemporal evolution of the quasiparticle wave packets with distance betwee wave peaks being about $2L_h$ are clearly exhibited in Fig. 1(c) for $|\psi_{\pm, 0 0 1}(x,t)|^2$ with $\phi'=1.8$ and the symmetrical evolution in Fig. 1(d) for $|\psi_{\pm, 1 0 1}(x,t)|^2$ with $\phi'=3$, in which no interchange of the quasiparticles occurs.

Note that at any moment $t_0$, the probabilities $P_{\pm}(t_0)= \int |\psi_{\pm, 0 0 1}(x,t_0)|^2dx$ of the electron being in different spin states are equal to the areas between the wave packet curves and the $x$ axes. The spatial distributions in Fig. 1(b) mean that the symmetrical kitten and ill kitten states have the same probability $P_{\pm}(5.0)=P_{\pm}(13.8)=\frac 1 2$. Obviously, for some times Fig. 1(a) has asymmetrical probability distributions to produce $P_+\ll P_-$ or $P_-\ll P_+$. In Fig. 1(c) and 1(d), the different density distributions are kept approximately for all times, meaning at any time $P_-(t)< P_+(t)$ in Fig. 1(c) and $P_-(t)\approx P_+(t)$ in Fig. 1(d). The asymmetrical superposition states similar to Fig. 1(c), of course, can be used to construct a new symmetrical superposition state according to the superposition principle of quantum states.

\emph{The electrically manipulated braiding operations} of the degenerate ground states can be achieved by a field-driven interchange of quasiparticles at an appropriate time interval $t\in [t_i, t_f]$, by using the time-evolution operator $U(t_f,t_i)=e^{-iH (t_f-t_i)}$ to act on the initial state $|\psi_{l n_u n_v}(t_i)\rangle$ which fits a stationary state of undriven system. We switch on the ac electric field at $t_i$ and switch off it at $t_f$, creating a quantum transition between the initial and final stationary states. Such stationary degenerate ground states will be demonstrated in the next section. For instance, starting at the initial time $t_i=0$, the operation time in Fig. 1(b) should be $t_f=5 {\omega^{-1}}$ for obtaining the ill kitten states, and the operation time for the quasiparticle interchange to yield the kitten state should be $t_f=13.8 {\omega^{-1}}$. By selecting other operation times in Fig. 1(a) or taking the parameters of Fig. 1(c), we can create the superposition states with larger probability in spin-up or spin-down state. The braiding operation based on the interchanges of the non-Abelian identical quasiparticles may be insensitive to noise and perturbations, because of the topologies of states. Such electrically controlling quasiparticle interchanges can be performed locally for any electron in an array of quantum-dot electrons \cite{Loss}. The operation times for different electrons can be selected to changes the state of the system in a way that depends only on the order of the exchanges.

\begin{figure}[htp]
\includegraphics[height=1.55in,width=1.7in]{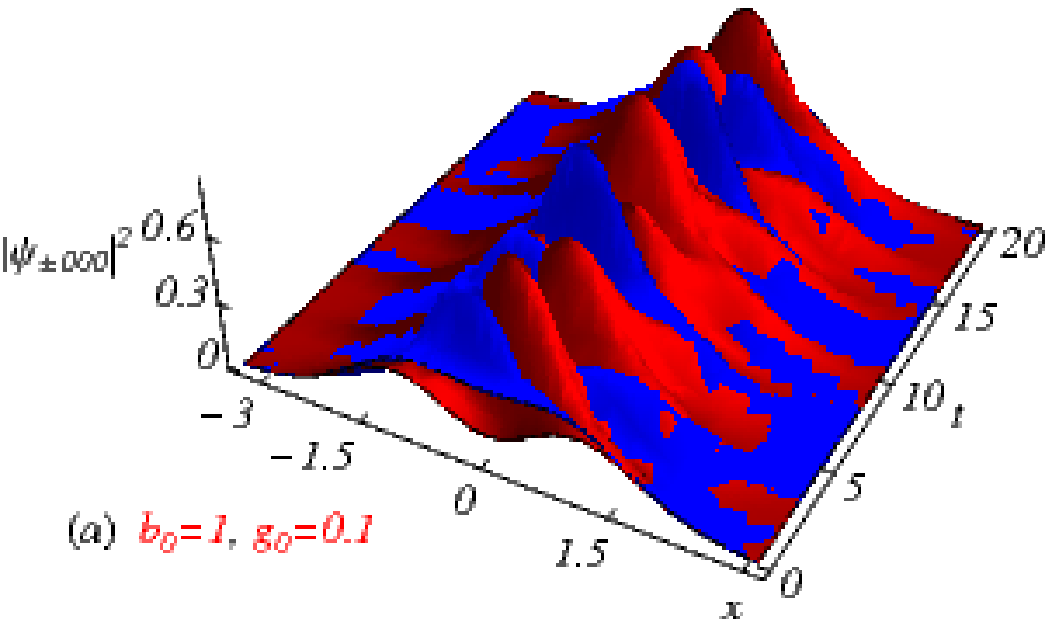}
\includegraphics[height=1.45in,width=1.65in]{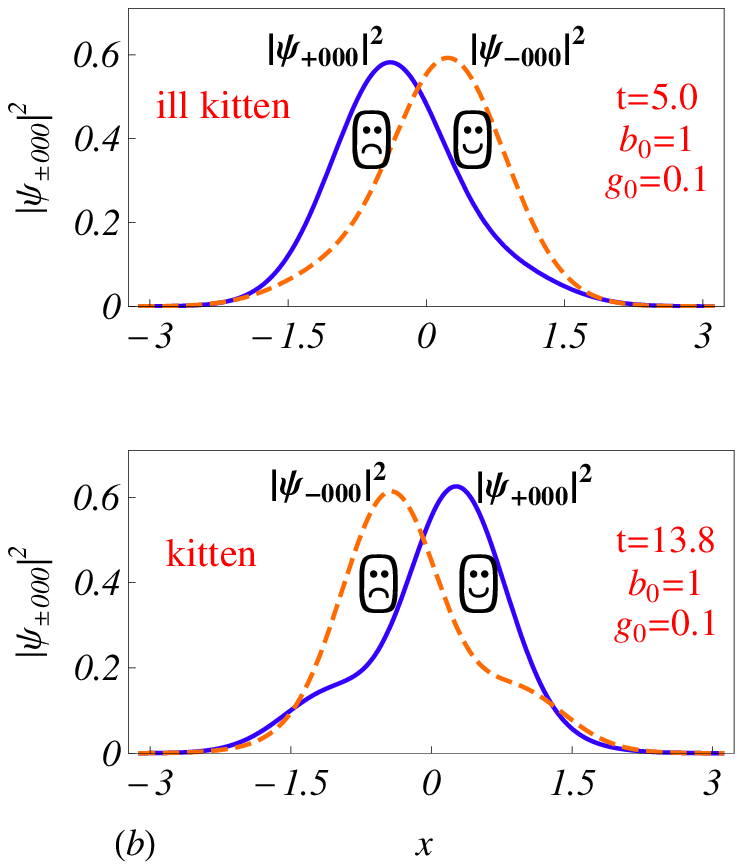}
\includegraphics[height=1.45in,width=1.67in]{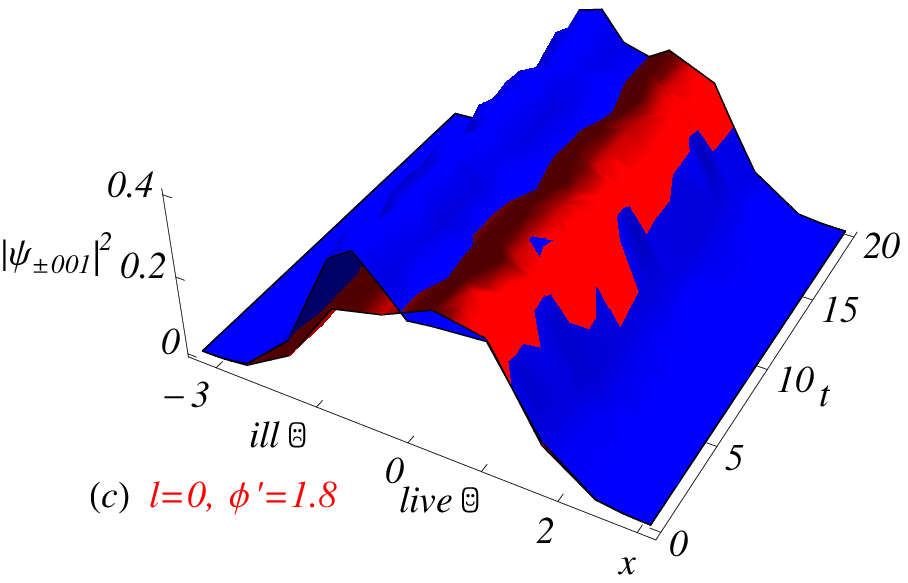}
\includegraphics[height=1.45in,width=1.67in]{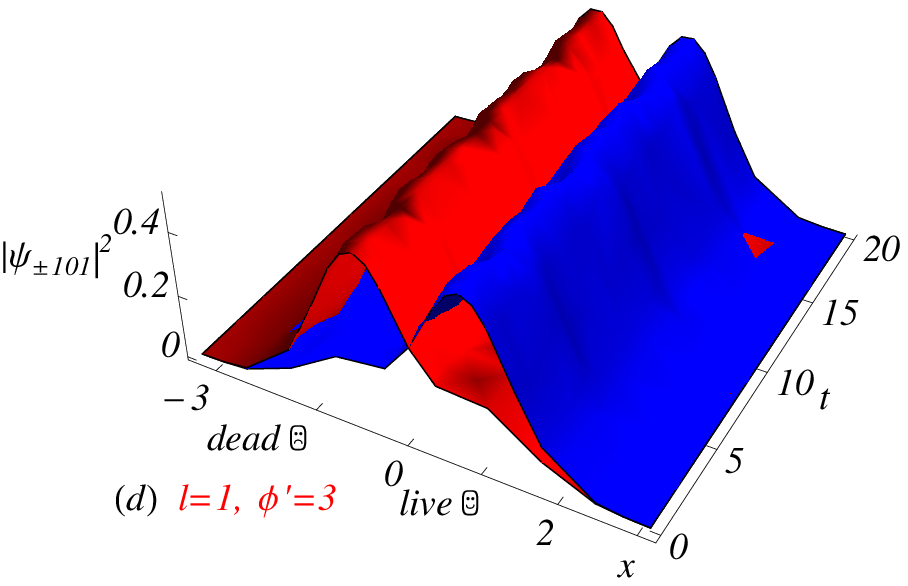}
\caption{(Color online)  Spatiotemporal evolutions of the quasiparticle wave packets in (a, b) for the motional ground states $\psi_{\pm, 0 0 0}(x,t)$ and (c, d) for the first excitation states $\psi_{\pm, l 0 1}(x,t)$. Hereafter, the deep colour (online blue) or solid curve is associated with sign ``$+$" and the light colour (online red) or dashed curve with sign ``$-$", and the right and left wave packets in a plot are always labelled by the live $\smiley{}$ and dead $\frownie{}$ respectively. When the packet $|\psi_{+, 0 0 0}(x,t)|^2$ is localized on right or left side, we called the superposition state the Schr\"odinger kitten state [e.g. the state at $t=13.8$ of Fig. 1(b)] or ``ill kitten" state [e.g. the state at $t=5$ of Fig. 1(b)]. The symmetrical probability densities in Fig. 1(b) mean the same occupying probability of different spin states, and the asymmetrical density distributions in Fig. 1(a) for the time interval $t\in(5.0, 13.8)$ and in Figs. 1(c) for all times mean the different occupying probabilities of the different spin states. All the quantities plotted in the figures of this paper are dimensionless.}
\end{figure}

In order to increase the oscillating amplitudes of the wave packets \cite{Monroe,Kienzler} for creating more Schr\"odinger kitten states, we can apply a $\pi/2$ pulse of Ramsey type experiment to rotate the state vector (9) to the form \cite{Gardiner}
$|\psi'_{l n_u n_v}\rangle=\frac{1}{\sqrt 2}[|\psi'_{+, l n_u n_v}\rangle|\uparrow \rangle+|\psi'_{-, l n_u n_v}\rangle|\downarrow \rangle]$
with $|\psi'_{\pm, l n_u n_v}\rangle=\frac{1}{\sqrt{2}}(|\psi_{+, l n_u n_v}\rangle\pm |\psi_{-, l n_u n_v}\rangle$. Thus the probability amplitudes of the electron being in $|\uparrow \rangle$ and $|\downarrow \rangle$ become $\psi'_{\pm, l n_u n_v}(x,t)=\langle x|\psi'_{\pm, l n_u n_v}\rangle=\frac {1}{\sqrt 2}[\psi_{+, l n_u n_v}(x,t)\pm \psi_{-, l n_u n_v}(x,t)]$. Then applications of Eq. (8) with $c_u=c_v=1$ give
\begin{eqnarray}
\psi'_{+, l n_u n_v}(x,t)&=& e^{\frac i 2(\alpha^2 t-l\pi)} \Big[f_{n_u}e^{-i(\alpha x+g_0 t)}\cos \frac {\phi}{2} \nonumber \\ &&-i f_{n_v}e^{i(\alpha x+g_0 t+l\pi)}\sin \frac{\phi}{2}\Big], \nonumber \\
\psi'_{-, l n_u n_v}(x,t)&=& e^{\frac i 2(\alpha^2 t-l\pi)} \Big[-i f_{n_u}e^{-i(\alpha x+g_0 t)}\sin \frac {\phi}{2} \nonumber \\ &&+ f_{n_v}e^{i(\alpha x+g_0 t+l\pi)}\cos \frac{\phi}{2}\Big].
\end{eqnarray}
The corresponding quasiparticle wave packets are described by the probability densities
\begin{eqnarray}
&&|\psi'_{+, l n_u n_v}(x,t)|^2 =\Big[R_{n_u}^2(x,t)\cos^2 \frac {\phi}{2}+R_{n_v}^2(x,t)\sin^2 \frac {\phi}{2}\Big] \nonumber\\
&+ & \sin \phi\ R_{n_u}R_{n_v}\sin [\Theta_{n_u}-\Theta_{n_v}-2(\alpha x+g_0 t)-l\pi],  \nonumber\\
&&|\psi'_{-, l n_u n_v}(x,t)|^2 = \Big[R_{n_u}^2(x,t)\sin^2 \frac {\phi}{2}+R_{n_v}^2(x,t)\cos^2 \frac {\phi}{2}\Big] \nonumber\\
&- & \sin \phi\ R_{n_u}R_{n_v}\sin [\Theta_{n_u}-\Theta_{n_v}-2(\alpha x+g_0 t)-l\pi]\ \ \
\end{eqnarray}
which obey the normalization requirement $|\psi'_{+, l n_u n_v}|^2+|\psi'_{-, l n_u n_v}|^2=R_{n_u}^2+R_{n_v}^2$. A careful calculation can prove that such a rotation keep the expectation value of energy (10) and the independence of energy on the parameters $\phi$ and $l$. Therefore, $|\psi'_{l n_u n_v}\rangle$ and $|\psi'_{l' n_u n_v}\rangle$ are also two degenerate states with different $|\psi'_{\pm,l n_u n_v}\rangle$ and $|\psi'_{\pm,l' n_u n_v}\rangle$, while for a fixed $l$ the magnetic angle transformation from $\theta=\phi$ to $\theta=\phi+\pi$ causes state transition between two degenerate states with exchange between $\psi'_{+, l n_u n_v}(x,t)$ and $\psi'_{-, l n_u n_v}(x,t)$, meaning the spin flip. We are interested in the exact ground state solution $\psi'_{\pm, l 00}(x,t)$ with $n_u=n_v=l=0$. Adopting the parameters $\alpha= 0.1, g_0= 0.5, b_0 = 3, \epsilon = 0.2$ and $\Omega= 2$, from Eq. (14) we plot the spatiotemporal evolutions of the quasiparticle wave packets $|\psi'_{\pm, 0 0 0}(x,t)|^2$ for Fig. 2(a) with $\phi= 0.1$ and Fig. 2(b) with $\phi= 0.1+\pi$. The governing initial states is a kitten state as Fig. 2(a) or an ill kitten state as Fig. 2(b). Comparing Fig. 2(a) with Fig. 1(a), we find that as the increase of the initial constant $b_0$ from $1$ to $3$, the maximal distance between wave packets is lengthened by a approximate factor $6$, while the constant $b_0$ can be selected by preparing the initial wave packets \cite{Monroe,Kienzler}. Along the line $x=0$, there are some points where the wave packets overlap periodically in time that enables periodic interchanges of wave packet positions. In the exchange process of two maximally separated wave packets, we can create many different kitten states with different distances between two wave packets, by switching off the ac field at different moments. Then we adjust the magnetic angle from $\theta=\phi=0.1$ to $\theta=\phi=0.1+\pi$ that brings wave packet exchanges between $|\psi'_{+, 0 0 0}(x,t)|^2$ and $|\psi'_{-, 0 0 0}(x,t)|^2$, as shown in Fig. 2(b). Such  \emph{The magnetically controlling quasiparticle interchanges} can be performed simultaneously in a wide range for an array of driven quantum-dot electrons.

\begin{figure}[htp]
\includegraphics[height=1.45in,width=1.73in]{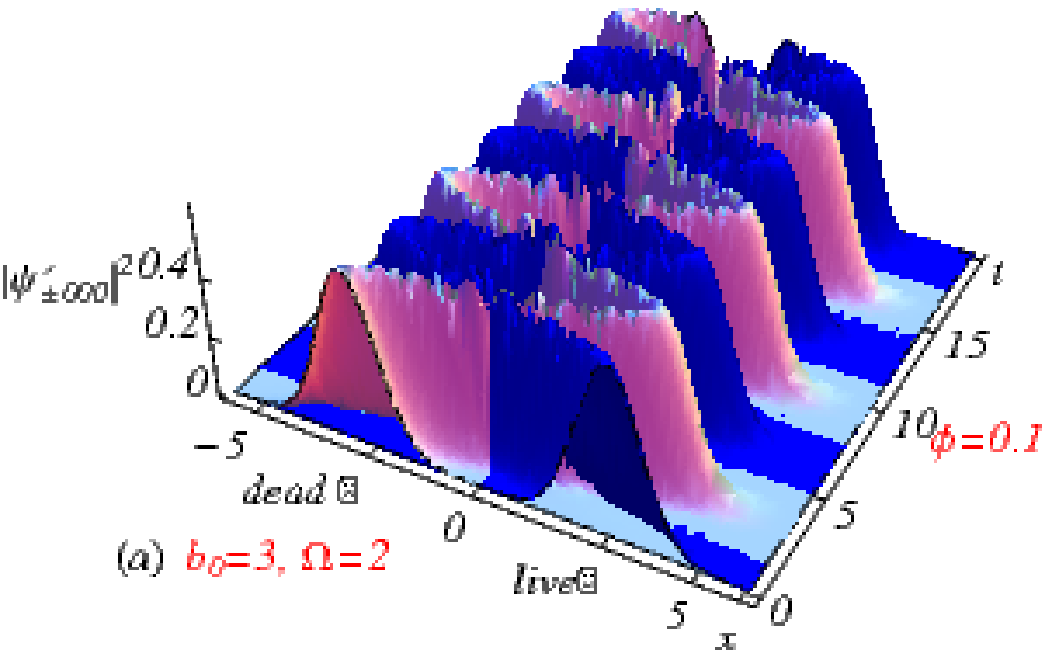}
\includegraphics[height=1.45in,width=1.62in]{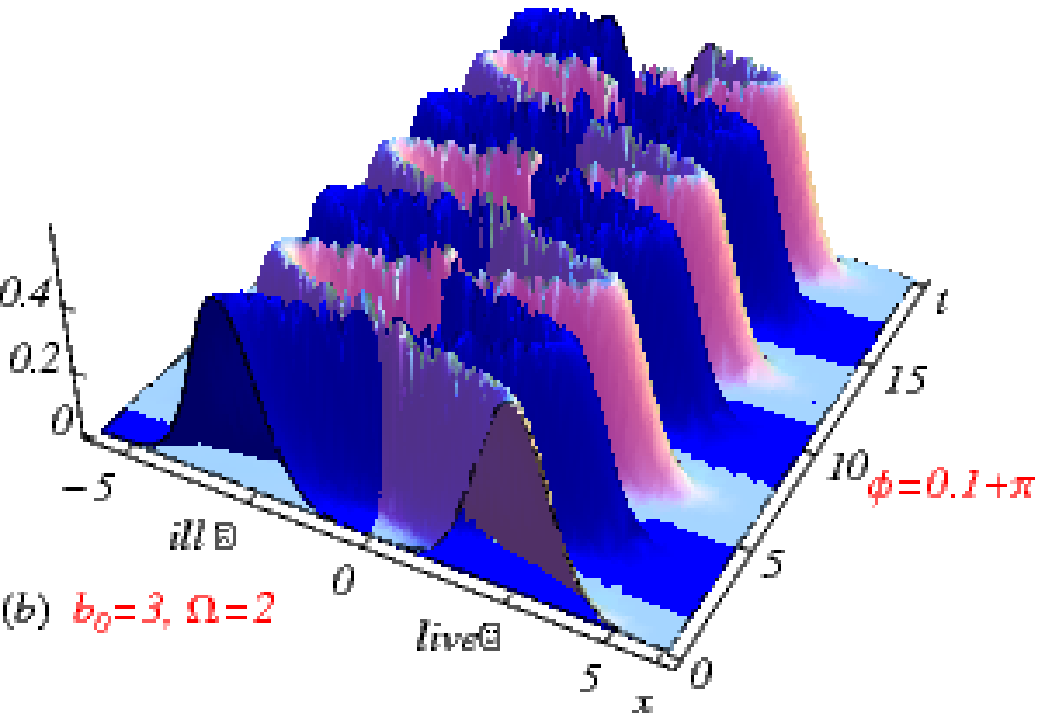}
\caption{(Color online)  Spatiotemporal evolutions of the quasiparticle wave packets $|\psi'_{\pm, 0 0 0}(x,t)|^2$ with a larger oscillating amplitude. The wave packet pairs periodically oscillate starting with a kitten state (a) or with an ill kitten state (b) respectively. In an oscillating period, the electron can go through many kitten and ill kitten states with different distances between wave packets, which can be extracted by electrical manipulation. While the transition from the state determined by the initial kitten state (a) to the state determined by the initial ill kitten state (b) is manipulated by varying the magnetic angle from $\theta=\phi=0.1$ to $\theta=\phi=0.1+\pi$.}
\end{figure}

\section{Coherent control of the stationary degenerate ground states}

Now we seek the stationary Majorana-like ground states of undriven case and focus in the coherent control of transitions between them by using the ac driving to perform the braiding operations based on interchanges of the quasiparticles. From Eq. (8) we know that the stationary ground states with $n_u=n_v=0$ cannot exist, because of the magnetic phase $\pm g_0t$. However, we will prove that under the \textbf{magnetic resonance} conditions \cite{Kato,Golovach,Rashba} $2 g_0=(n_v-n_u)(\hbar\omega)$, Eq. (8) becomes the stationary states with time-independent norms.
In fact, in the case $\epsilon=0$, the initial constant set $S_{u(v)}=[\gamma_{u(v)}(0), b_{u(v) 1}(0), b_{u(v) 2}(0), A_1, A_2, B_1, B_2]=[0,0,0,A,A,0,-\pi/2]$ makes the functions $f_{u(v)}$ of Eq. (7) the usual eigenstates of a harmonic oscillator. Taking a minimal resonance magnetic field with $g_0=(n_v-n_u)/2=1/2$ and inserting it into Eqs. (8) and (9) produces a set of \emph{stationary Schr\"odinger kitten states}, which contains the degenerate ground states $|\psi_{l 0 1}\rangle$ with $n_u=0,n_v=1$ and the motional states of Eq. (12) as
\begin{eqnarray}\label{dy3}
\psi_{\pm, l 0 1}= \frac{ e^{\frac i 2(\alpha^2 t-2t\mp \phi-l\pi)-i\alpha x-\frac{1}{2}
x^2}}{\pi^{\frac 1 4}\sqrt 2} (c_u \pm c_v \sqrt 2 x e^{i2\alpha x+il\pi}) \ \
\end{eqnarray}
for $l=$ even and odd numbers respectively. In the case $c_u=1, c_v=e^{i\phi'}$, Eq. (10) means that Eq. (15) becomes a set of degenerate ground states.

Writing the phases of Eq. (15) as $\Phi_{\pm}= \arg [\psi_{\pm, l 0 1}(x,t)]$, we have the time-independent phase gradients which contain some singular points for some values of the parameters $\alpha$ and $\phi'$. These singular points are similar to the vertex cores at which the densities vanish and phases hop for the motional states. To simplify, as an example, we consider only the ground states with the phase gradients
\begin{eqnarray}\label{dy3}
\Phi_{\pm,x}&=& \frac{4\alpha x^2 \pm \sqrt 2 [2\alpha x \cos (2\alpha x+\phi')+ \sin(2\alpha x+\phi')]}{1+2x^2\pm 2\sqrt 2 \ x\cos (2\alpha x+\phi')} \nonumber \\&&-\alpha. \ \
\end{eqnarray}
Zero points of the denominator imply that for $\alpha=\frac {1}{\sqrt 2}(n\pi-\phi')>0$ with $n=\pm1, \pm2,...$, the \emph{singular points} of $\Phi_{\pm,x}(x)$ are $x_{\pm}=\pm \frac {1}{ \sqrt 2}$, respectively. The required SOC is adjusted by the phase difference $\phi'$, and a usual zero phase difference corresponds to stronger SOC. To see the 1D topological property of the degenerate ground states, we can employ the analytic prolongation \cite{Goldstein} $\Phi_{\pm,x}(z)$ on the complex plan $z=x+iy$ to construct the circulation integrals $\oint_{\Gamma_{\pm}} \Phi_{\pm,x}(z) dz =2\pi i\times\text{res} \Phi_{\pm,x}(z_{\pm})=2N\pi$ for the topological charges $N=0,\pm 1,\pm 2,...$. Here $\Gamma_{\pm}$ are closed trajectories
enclosing the poles $z_{\pm}=x_{\pm}$ and the res denotes the residues at the poles. Various topologically equivalent closed trajectories are allowable for any one of the above circulation integral that reminds us the emergence of the similar topologies with planar vortices.

The kitten states $|\psi_{l 0 1}\rangle$ with motional states of Eq. (15) contain the maximally entangled state with the probability $P_{\pm, l 01}=\frac 1 2 \int |\psi_{\pm, l 01}(x,t)|^2 dx=\frac 1 2$ for different values of $\alpha$ and/or $\phi'$. The degenerate first excitation state reads $|\psi_{l 1 2}\rangle$ with $n_u=1,n_v=2$ for different $l$ values. The corresponding eigenenergies are given by Eq. (10) as $E_{01}=(1-\frac 1 2\alpha^2)(\hbar\omega)$ and $E_{12}=(2-\frac 1 2\alpha^2)(\hbar\omega)$. The energy gap $\Delta E=E_{12}-E_{01}=1(\hbar\omega)$ is relatively great compared to the perturbation level difference in Ref. \cite{RLi}. The large energy gap may be important for performing the fault tolerant topological quantum computation \cite{Stern}.

It is easy to create a usual quantum transition from ground state $|\psi_{l 0 1}\rangle$ to excitation state $|\psi_{l 1 2}\rangle$ by using a laser with resonance frequency to match the level difference $\Delta E$. However, the topological quantum computation needs the braiding operations to the different degenerate ground states distinguished by the parameters $l$ and $\phi'$. Although Eq. (15) contains many different ground states with symmetrical or asymmetrical wave packets, we here consider only a simple example. For the SOC strength value $\alpha=0.1$ from Eq. (15) with $c_u=1, c_v=e^{i\phi'}=e^{i 3}$ we plot the density wave packets $|\psi_{\pm, 0 0 1}|^2$ and $|\psi_{\pm, 1 0 1}|^2$, as shown in Fig. 3. According to the definition of a kitten state, Fig. 3(a) with $l=0$ is associated with an ill kitten state and Fig. 3(b) with $l=1$ corresponds to its degenerate kitten state. The distance between two packets is in units of $\sqrt{\hbar/(m_e\omega)}$ adjusted by the gate voltages on the static electric gates. In Fig. 3(a) and 3(b) we also show that for a fixed SOC strength the density wave packets obey $|\psi_{+, 0 0 1}|^2=|\psi_{-, 1 0 1}|^2$ and $|\psi_{-, 0 0 1}|^2=|\psi_{+, 1 0 1}|^2$, and the former has the zero density nodes $x_+=\frac {1}{\sqrt 2}$. This means that the interchanges of the wave packets with $l=0$ to those with $l=1$ implies quantum transitions between the degenerate ground states $|\psi_{0 0 1}\rangle$ and $|\psi_{1 0 1}\rangle$. In Fig. 3(c) and 3(d), we show that the height, width, symmetry and rich deformations of the wave-packet pairs and the distance between two packets of any pair are tuned by the SOC strength and the phase difference.

\emph{Controlling transitions between stationary ground states}. As shown in Figs. 1(a) and 2(a), in an oscillating period of the wave packets, the electron can experience many kitten and ill kitten states with different heights, widths, symmetry of wave packets and distances between them. Therefore, starting with any one of the states in Fig. 3 with the same parameters, we can switch on the ac field to drive the wave packets, then switch off the driving at a suitable time $t_f$ for transferring the state to another of Fig. 3. The accumulated phase in the driving process can be fitted by the phases of complex numbers $c_u$ and $c_v$ in Eq. (15). Combining the electric and magnetic manipulation shown in Fig. 1, we can control the quantum transitions between different pairs of the stationary degenerate ground states. In addition, we can easily illustrate that the transitions between the degenerate ground states are robust and insensitive to various perturbations. For instance, when we vary the parameters in the ranges $\alpha=0.1\pm 0.1$ and $\phi'=3\pm 0.1$, the produced wave packets have only very small deformations from Fig. 3(a) and 3(b). Moreover, when the operation time $t_f$ is taken in a small interval, the extracted wave packets from Figs. 1 and 2 are similar. Such manipulations may be useful for braiding the degenerate non-Abelian quasiparticles to realize topological quantum gates.

\begin{figure}
\begin{center}
\includegraphics[height=1.2in,width=1.6in]{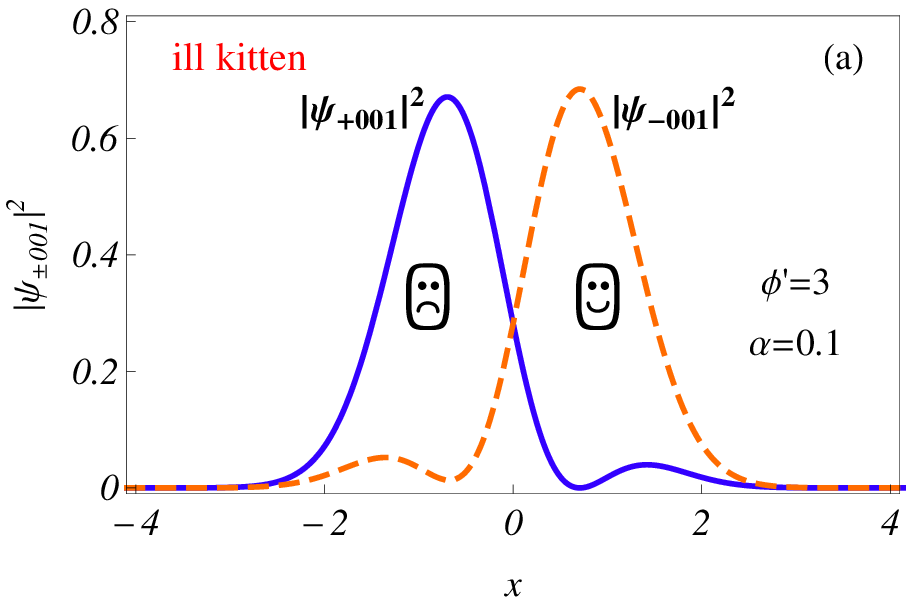}
\includegraphics[height=1.2in,width=1.6in]{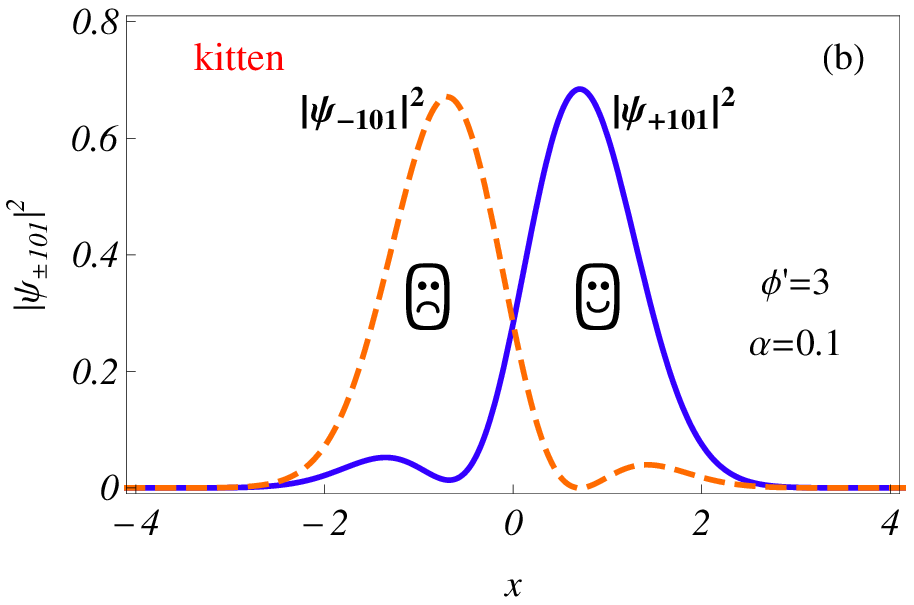}
\includegraphics[height=1.2in,width=1.6in]{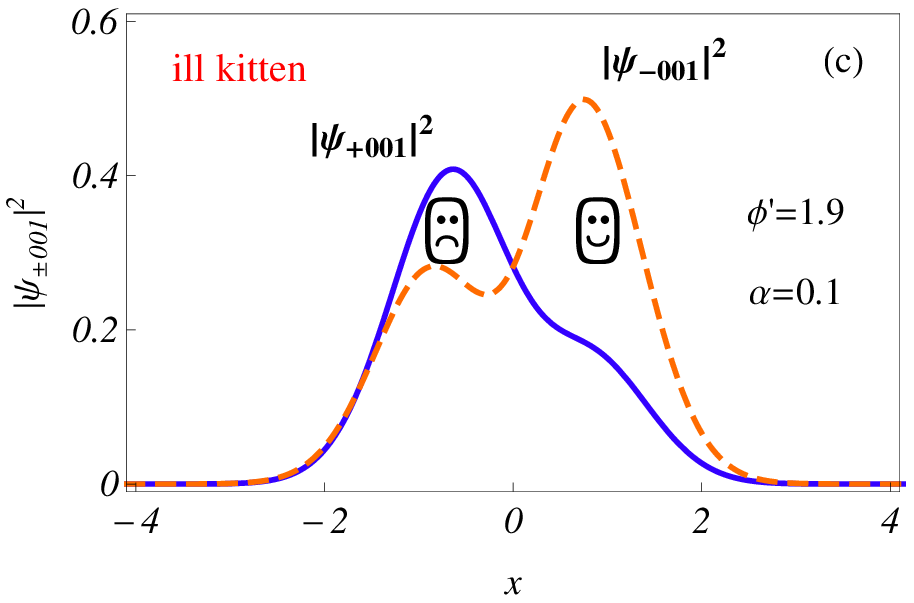}
\includegraphics[height=1.2in,width=1.6in]{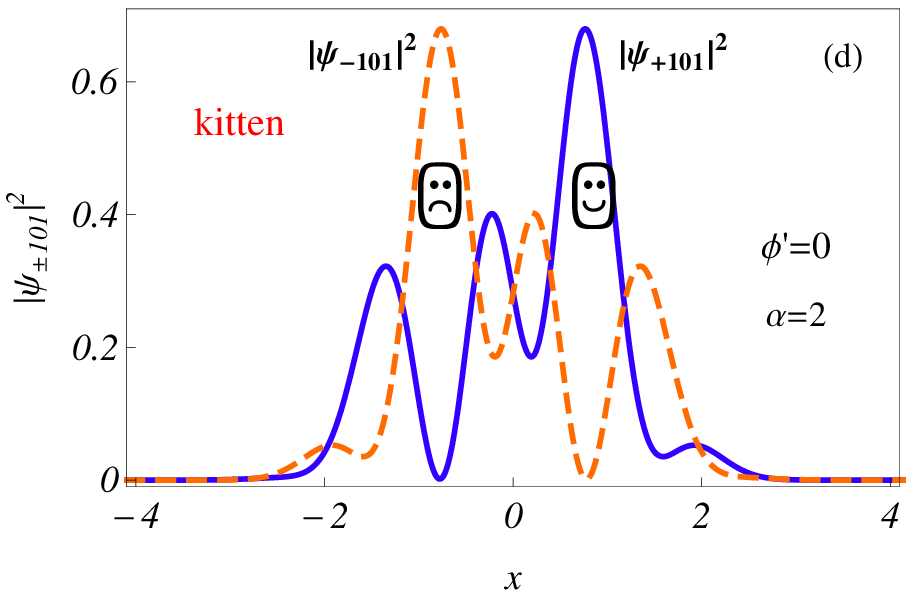}
\caption{(Color online) The spatial distributions of wave packets associated with four stationary degenerate ground states: (a) an ill kitten state with $l=0, \phi'=3, \alpha=0.1$, (b) a kitten state with $l=1, \phi'=3, \alpha=0.1$, (c) an asymmetric ill kitten state with $l=0, \phi'=1.9, \alpha=0.1$, and (d) a deformed kitten state with $l=1, \phi'=0, \alpha=2$. Quantum transition between these degenerate ground states can be electrically manipulated.}
\end{center}
\end{figure}

\section{Conclusions and discussions}

We have investigated a single spin-orbit coupled quantum-dot electron, subject to an ac electric field. Under the magnetism- and SOC-dependent phase-locked condition, we derive a set of complete solutions of Schr\"odinger equation with arbitrary constants adjusted by the initial conditions, which describes a complete set of Schr\"odinger kitten states and contains some novel degenerate ground states with oscillating wave packets. In the undriven case, pairs of stationary wave packets of degenerate ground states are constructed. The degeneracy is not based on simple symmetry consideration and is topological thereby. We identify such wave packets as Majorana-like quasiparticles and demonstrate that they obey non-Abelian statistics and behave as electroneutrality without Coulomb interaction between them. The braiding operations based on the interchanges of the degenerate non-Abelian quasiparticles with one wavepacket going through another are shown, which shift the system between different ground states. The exact results can be directly extended to the 2D quantum-dot-electron system with SOC Hamiltonian \cite{Golovach} $H_{SO}=(\alpha_R\sigma_y-\alpha_D\sigma_x) p_x+(\alpha_D\sigma_y-\alpha_R\sigma_x) p_y$ for the special case $\alpha_D=\alpha_R$ in which the 2D Hamiltonian possesses invariance in exchange between $x$ and $y$ that will results in novel planar vortex states and 2D strip states \cite{Ozawa}. Treating the exact solutions as leading-order solutions, the obtained results could be directly extended to an array of electrons separated from each other by different 2D quantum dots with weak neighboring coupling as perturbation for topological quantum computation.  The braiding operations based on the interchanges of the non-Abelian identical quasiparticles may be insensitive to perturbations and weak noise from the environment. The operation can be performed individually for any one of the quantum-dot electrons \cite{Loss}. The operation times for different electrons can be selected to changes the state of the system in a way that depends only on the order of the exchanges. The quantum operations can be performed adiabatically by reducing the driving frequency or ultrafast by applying an array of ultrashort laser pules to replace the periodic driving in Eq. (1) \cite{Hai2,Mizrahi}.

In a tight-binding approximate system, the Majorana quasiparticles are localized at some spatially separated positions with a certain probability at any time and their interchange is accompanied by the perfectly predictable time evolution of their wavefunctions in Hilbert space. Differing from that, our Majorana kitten-particles periodically separate and overlap in continuous coordinate space and the spatiotemporal evolution of the exact solution governs the quasiparticle interchange with non-Abelian statistics. Our results exactly reveal the coherent control of a qubit in a 1D solid-state electronic system, which could be fundamental important for designing solid-state quantum circuits \cite{Loss}.
To do useful computations, one needs to create many Majorana-like particles, and to develop the ability to move their spins \cite{Wilcze}.
We can propose an implementation of qubit gates for topological quantum computation using the spin states of coupled single-electron quantum dots. Desired braiding operations are effected by the gating of the tunneling barrier between neighboring dots. Arrays of quantum dots of the type developed by D. Loss and D. P. DiVincenzo \cite{Loss} could support our scheme.

\bf Acknowledgments \rm
This work was supported by the NNSF of China under Grant Nos. 11475060 and 11204077.

\end{document}